**TITLE**

Spared cognitive and behavioral functions prior to epilepsy onset in a rat model of subcortical band heteropia

**AUTHOR NAMES AND AFFILIATIONS**

Fanny Sandrine Martineau[a], Lauriane Fournier[a], Emmanuelle Buhler[a], Françoise Watrin[a], Francesca Sargolini[b], Jean-Bernard Manent[a], Bruno Poucet[b] and Alfonso Represa[a]

[a]INMED, Aix-Marseille University, INSERM U1249, Marseille 13273 CEDEX 09, France

[b]LNC – Fédération de recherche 3C, Aix-Marseille University, CNRS UMR7291, Marseille 13331 CEDEX 03, France

**CORRESPONDING AUTHOR**

Alfonso Represa, INMED, Parc scientifique de Luminy, 13273 Marseille CEDEX 09 (France)

alfonso.represa@inserm.fr

**ABSTRACT**

Subcortical band heterotopia (SBH), also known as doublecortex syndrome, is a malformation of cortical development resulting from mutations in the *doublecortin* gene (*DCX*). It is characterized by a lack of migration of cortical neurons that accumulate in the white matter forming a heterotopic band. Patients with SBH may present mild to moderate intellectual disability as well as epilepsy. The SBH condition can be modeled in rats by *in utero* knockdown (KD) of Dcx. The affected cells form an SBH reminiscent of that observed in human patients and the animals develop a chronic epileptic condition in adulthood. Here, we investigated if the presence of an SBH is sufficient to induce cognitive impairment in juvenile Dcx-KD rats, before the onset of epilepsy. Using a wide range of behavioral tests, we found that the presence of SBH did not appear to affect motor control or somatosensory processing. In addition, cognitive abilities such as learning, short-term and long-term memory, were normal in pre-epileptic Dcx-KD rats. We suggest that the SBH presence is not sufficient to impair these behavioral functions.

**ACKNOWLEDGEMENTS**

We thank the animal facility (PPGI, INMED, Marseille), Dr Stephane Gaillard, Irene Sanchez Brualla and Dr Frederic Brocard for their help with the Von Frey hair test, Robert Martinez and Dr Ludovic Petit for their help building and lighting the texture discrimination box and Dr Bruno Bontempi for his help with the social transmission of food preference test. We thank Dr Etienne Save for critical discussions and comments on the article.

**FUNDING SOURCES**

This work was supported by The French National Agency for Research [grant number #19012012 SAMENTA], the European Community 7th Framework programs [grant number Health-F2-602531-2013, DESIRE], La Fondation pour la Recherche Médicale [grant number #FDT20160435216].

**CONFLICT OF INTEREST**

The authors declare no competing financial interests.

1. INTRODUCTION

Subcortical band heterotopia (SBH) is a malformation of cortical development arising from migration failure during corticogenesis. In this malformation, some newborn neurons fail to migrate to the cortical plate and aggregate in the white matter below the future cortex. Accordingly, on MRI images, SBH can be seen as a band of grey matter located in the subcortical white matter (Barkovich et al., 1989; Watrin et al., 2015). Other neurological abnormalities can be associated with SBH, such as enlarged ventricles, pachygyria or simplified gyration pattern. In most cases, the brain malformation is identified after epileptic seizures onset, usually within the first decade of the patient's life (Tanaka and Gleeson, 2007). About 85% to 96% of SBH patients suffer from epilepsy, with a high proportion of drug-resistant cases (Bahi-Buisson et al., 2013; Tanaka and Gleeson, 2007). In addition, SBH is often associated with developmental delay and mild to moderate intellectual disability. Clinical evaluations show heterogeneous impairments including delayed motor development, reduced processing speed, learning deficiency, low IQ and memory deficits (Bahi-Buisson et al., 2013; Barkovich et al., 1994; Jacobs et al., 2001; Janzen et al., 2004; Tanaka and Gleeson, 2007). SBH is typically encountered in female patients because an X-linked gene, doublecortin (*DCX*), is responsible for most instances of this malformation (Des Portes et al., 1998; Gleeson et al., 2000, 1998; Matsumoto et al., 2001). Males with *DCX* mutations usually suffer from lissencephaly, or "smooth brain syndrome" but some cases of SBH caused by mosaic mutations have also been reported (D'Agostino et al., 2002; Gleeson et al., 2000; Guerrini et al., 2003; Janzen et al., 2004; Poirier et al., 2013). In rats, mosaic knockdown of Dcx induced by *in utero* electroporation reproduces the migration defect seen in patients and leads to the formation of a SBH in the white matter (Bai et al., 2003). Around 6 months of age, electroporated animals start displaying spontaneous epileptic seizures akin to those seen in patients (Lapray et al., 2010). Over the years, the Dcx knockdown (Dcx-KD) rat has helped unravelling part of the pathophysiological mechanisms of SBH (Lapray et al., 2010; Manent et al., 2009; Ramos et al., 2006). Thanks to this model, the understanding of network events leading to epilepsy associated with SBH has grown tremendously (Ackman et al., 2009; Petit et al., 2014). However, to this day, other comorbidities associated with SBH such as motor and cognitive impairments have been scarcely investigated in animal models. Here, we tested the behavior of 2-month old Dcx-KD rats, when animals did not yet display any epileptic feature (Lapray et al., 2010), to check if the presence of SBH induces motor, somatosensory or cognitive impairments. Despite the presence of SBH, basic motor control and processing of somatosensory information were found to be normal in these animals. In addition, Dcx-KD rats showed that their ability to learn remained unaltered by the presence of the SBH in no less than 4 tasks. Finally, both short-term memory and long-term memory seem spared in Dcx-KD rats. Our data suggest that the presence of a SBH in itself is not enough to induce cognitive impairment.

2. RESULTS

In order to assess how the presence of SBH affects motor control and cognitive functions, we reproduced this malformation in rat by knocking down Dcx *in utero*. While brains electroporated with a short hairpin RNA (shRNA) targeting Dcx display a SBH, brains electroporated with a control mismatch shRNA present no such alteration (Figure 1A).

*2.1. Anxiety and motor functions*

First, we controlled that anxiety was unchanged in rats with SBH by using the elevated plus maze (Figure 1B-C). During this task, Dcx-KD rats spent as much time in the more stressful open arms as mismatch rats (31.40 ± 11.26 vs 34.50 ± 8.77 seconds; Mann-Whitney test, p = 0.6277, n = 5-6 / condition). Moreover, the total locomotor activity was similar in both conditions (65.16 ± 6.95 vs 65.46 ± 3.73 meters; Mann-Whitney test, p = 0.8745, n = 5-6 / condition). These results indicate that Dcx-KD rats do not display any elevated anxiety that could interfere with other tests.

Because some SBH patients show a delayed motor development, we proceeded to test motor functions of Dcx-KD rats. Animals were trained on the rotarod for 10 sessions each day during 3 days and their latency to fall was measured for each session (Figure 1D). Performances of mismatch and Dcx-KD rats were not significantly different for this task, no matter the rotating speed considered (5 rpm: 140.80 ± 21.33 vs 126.60 ± 13.64 seconds, 10 rpm: 153.70 ± 16.86 vs 113.60 ± 25.07 seconds, 15 rpm: 109 ± 32.58 vs 69.77 ± 25.46 seconds; Mann-Whitney tests; p = 0.4589, p = 0.3268 and p = 0.329 for 5 rpm, 10 rpm and 15 rpm respectively; n = 5-6 / condition). The rotarod task was followed by a beam walking test to check whether the malformation affects finer motor skills. In this task, mismatch and Dcx-KD rats performed similarly regarding the latency to fall (139.30 ± 28.63 vs 125.80 ± 35.50 seconds; Mann Whitney test; p = 0.9242), path length (1079 ± 200.44 vs 775.20 ± 190.04 centimeters; Mann-Whitney test, p = 0.329) and average speed (8.10 ± 0.47 vs 8.01 ± 1.66 centimeters / second; Mann-Whitney test, p = 0.5281) (n = 5-6 / condition) (Figure 1E-G;). Overall, these data indicate that basic motor functions seem preserved in Dcx-KD rats.

2.2. *Somatosensory functions*

In this animal model, we target the right side during the *in utero* electroporation so the SBH forms in the right hemisphere, below the somatosensory cortex (Petit et al., 2014). Accordingly, we sought to evaluate if the malformation impacts somatosensory processing. First, we probed the tactile sensitivity of the rats' hind paws using the Von Frey hair test (Figure 2A). Mechanical 50% thresholds were not significantly different between the right and left hind paws of animals (Mismatch: 5.79 ± 0.65 vs 6.04 ± 1.14; Dcx-KD: 6.80 ± 1.17 vs 5.09 ± 0.67; ordinary 2-way ANOVA, paw effect, $F_{1,36}$ = 0.5958, p = 0.4452, n = 10 / condition) and were similar between both groups (ordinary 2-way ANOVA, paw x group interaction, $F_{1,36}$ = 1.096, p = 0.3021, n = 10 / condition).

Second, the adhesive removal test evaluated the haptic sense of the front limbs (Figure 2B-C). Latency to contact all adhesive dots was similar between groups (ordinary 2-way ANOVA, limb x group interaction, $F_{1,36}$ = 0.3192, p = 0.5756, n = 10 / condition) and between both limbs within each group (Mismatch: 22.60 ± 6.15 vs 16.50 ± 3.21; Dcx-KD = 16.50 ± 2.49 vs 15.10 ± 3.87; ordinary 2-way ANOVA, limb effect, $F_{1,36}$ = 0.8128, p = 0.3733, n = 10 / condition). Furthermore, 80% (8/10) of Dcx-KD rats first contacted a dot on their right side suggesting a lack of sensitivity on their left side, but this result did not reach significance (binomial tests, p = 0.7539 (mismatch) and p = 0.1094 (Dcx-KD)).

Third, a gap crossing test was used to check vibrissae-dependent object perception (Figure 2D-E). On average, Dcx-KD rats could cross the same maximal distance as mismatch rats (13.30 ± 0.47 vs 13.10 ± 0.23 cm; Mann-Whitney test, p = 0.4663). Because the SBH is unilateral, there was a chance rats could compensate a deficit of their left vibrissae caused by the malformation with their right vibrissae. Therefore, we clipped the right vibrissae of all rats and tested them again on the gap crossing test. Again, rats from both groups could cross the same maximal distance, suggesting that vibrissae-dependent object detection is not impaired in Dcx-KD rats (14.40 ± 1.02 vs 13.90 ± 0.67 cm; Mann-Whitney test, p = 0.5631). Finally, we designed a new texture discrimination test to assess the integration of sensory information in a complex task (Figure 2F). We used sandpapers of 2 different grits to make the compartments of a testing box distinguishable in the dark. This box was paired with a conditioned place preference protocol, where one compartment became associated with aversive strobe light. The test was initially divided in 3 sessions: (1) animals were allowed to freely explore the environment; (2) animals were randomly locked in a compartment and underwent aversive conditioning; (3) animals were allowed to move freely in the whole box again.

To properly carry out this test, we first had to check that Dcx-KD rats did not display any conditioning deficits. To this end, the sandpaper was replaced by proximal spatial cues to distinguish both compartments and the test was run in a lit environment ("Phase I"). As seen in Figure 2G, the aversive conditioning succeeded as animals preferentially spent their time in the compartment where they were not dazzled by strobe light (ordinary 2-way ANOVA, compartment effect, $F_{1,32}$ = 32.47, p < 0.0001, n = 10 / condition) and mismatch and Dcx-KD rats explored the testing box in a similar way (Devaluated compartment: -37.67 ± 8.03 vs -47.70 ± 5.22; Non-devaluated compartment: 48.75 ± 15.69 vs 37.44 ± 22.37; ordinary 2-way ANOVA, compartment x group interaction, $F_{1,32}$ = 0.0018, p = 0.9666, n = 10 / condition).

During the testing phase ("Phase II"), compartments were covered with different sandpapers and rats were put in the dark to repeat the experiment (Figure 2H). Although the results were less striking, animals from both groups preferentially explored the non-devaluated compartment in session 6 (ordinary 2-way ANOVA, compartment effect, $F_{1,34}$ = 6.253, p = 0.0174, n = 10 / condition) without any significant difference between mismatch and Dcx-KD rats (Devaluated compartment: -10.333 ± 14.738 vs -30.2 ± 10.444; Non-devaluated compartment: 42.44 ±30.90 vs 17.60 ± 17.22; ordinary 2-way ANOVA, compartment x group interaction, $F_{1,34}$ = 0.0153, p = 0.9022, n = 10 / condition). Therefore, Dcx-KD rats could properly use somatosensory information to finely differentiate closely-related textures.

### 2.3. *Learning and memory functions*

We went on to test cognitive functions with a focus on learning and memory. To this end, we used the Morris water maze task and trained the animals during 8 sessions of 4 trials. As shown in Figures 3A-C, the escape latency and path length decreased across sessions (repeated measure 2-way ANOVA, session effect, $F_{7,63}$ = 4.269, p = 0.0006 and $F_{7,63}$ = 5.988, p = 0.0001, respectively, n = 6), showing that Dcx-KD rats were able to learn this spatial navigation task. Moreover, their performance was similar to that of mismatch rats indicating no impairment of their learning abilities (repeated measure 2-way ANOVA, session x group interaction, $F_{7,63}$ = 0.2997, p = 0.9514 (escape latency); $F_{7,63}$ = 0.5466, p = 0.7958 (path length) , n = 5-6 / condition). When training ended, we proceeded to test long-term memory of Dcx-KD rats by carrying out probe tests, consisting of 1 minute spent in the pool without the platform, at 2 different time points (24 hours and 10 days after the end of training) (Figure 3D-I). During both probe tests, Dcx-KD rats spent more time in the target quadrant than in the opposite quadrant (24 hours: 31.72 ± 1.70 vs 4.27 ± 1.61 seconds; 10 days: 18.72 ± 2.40 vs 8.75 ± 1.55 seconds), suggesting that they remembered the platform position during the training (ordinary 2-way ANOVA, quadrant effect, $F_{1,18}$ = 222.3, p < 0.0001 (24 hours); $F_{1,18}$ = 39.15, p < 0.0001 (10 days), n = 6). Dcx-KD rats also swam a greater distance in the target quadrant than in the opposite one (24 hours: 711.30 ± 50.21 vs 107.70 ± 35.69 centimeters; 10 days: 427.32 ± 54.85 vs 253 ± 39.58 centimeters) which seemed to confirm

this hypothesis (ordinary 2-way ANOVA, quadrant effect, $F_{1,18}$ = 114.8, p < 0.0001 (24 hours); $F_{1,18}$ = 22.88, p < 0.0001 (10 days), n = 6). In addition their performance was similar to that of mismatch for the time spent in each part of the pool (24 hours: 32.74 ± 2.67 vs 31.72 ± 1.70 seconds (target quadrant), 3.32 ± 1.58 vs 4.27 ± 1.61 seconds (opposite quadrant), ordinary 2-way ANOVA, quadrant x group interaction, $F_{1,18}$ = 0.2668, p = 0.6118; 10 days: 21.46 ± 2.83 vs 18.72 ± 2.40 seconds (target quadrant), 5.26 ± 1.13 vs 8.75 ± 1.55 seconds (opposite quadrant), ordinary 2-way ANOVA, quadrant x group interaction, $F_{1,18}$ = 2.222, p = 0.1534; n = 5-6 / condition). The path length was also similar in both groups (24 hours: 639.04 ± 74.17 vs 711.30 ± 50.21 centimeters (target quadrant), 106.26 ± 51.12 vs 107.70 ± 35.69 centimeters (opposite quadrant), ordinary 2-way ANOVA, quadrant x group interaction, $F_{1,18}$ = 0.4458, p = 0.5128; 10 days: 474.74 ± 68.70 vs 427.32 ± 54.85 centimeters (target quadrant), 147.98 ± 43.47 vs 253 ± 39.58 centimeters (opposite quadrant)), ordinary 2-way ANOVA, quadrant x group interaction, $F_{1,18}$ = 2.117, p = 0.1629, n = 5-6 / condition). Finally, although the error score, which reflects search dispersion, was similar between both groups at 24 hours post-training (Figure 3F; 34.90 ± 2.29 vs 36.17 ± 1.53; Mann-Whitney test, p = 0.99, n = 5-6 / condition), Dcx-KD rats scored a higher error than mismatch when tested 10 days later (Figure 3I; 36.74 ± 1.73 vs 44.60 ± 2.32; Mann-Whitney test, p = 0.0303; n = 5-6 / condition). This result suggests that Dcx-KD rats might exhibit a slight memory deficit specifically after extended periods of time.

To further investigate this hypothesis, we carried out the social transmission of food preference test, a paradigm designed to assess non-spatial memory after long delays (Figure 3J and (Bessières et al., 2017)). The test relies on a social behavior occurring in the wild where rats use the breath of congeners to determine safe food sources (Galef and Wingmore, 1983). In brief, healthy non-electroporated rats (demonstrators) ate cumin-flavored food before interacting with mismatch and Dcx-KD rats (observers). Because of this interaction, cumin flavor was then supposed to be considered as safe for mismatch and Dcx-KD rats. Both experimental groups were presented with cumin- and thyme-flavored foods 30 days later and, as expected for normal rats, Dcx-KD animals preferentially ingested cumin-flavored food (75%) over the usually more appealing thyme-flavored one (25%) (Bessières et al., 2017 and Figure 3K; ordinary 2-way ANOVA: flavor effect, $F_{1,36}$ = 3.237, p = 0.0084, n = 10). Their performance did not significantly differ from that of mismatch rats (ordinary 2-way ANOVA: flavor x group interaction, $F_{1,36}$ = 3.237, p = 0.0804, n = 10 / condition). Overall, considering this set of data, it seems that learning and long-term memory are essentially normal in Dcx-KD rats.

The experiments described above showed a preserved long-term memory in Dcx-KD rats but did not exclude a potential impairment of short-term memory. Different aspects of short-term memory were therefore tested, including working memory for which patients reports are contradictory (Jacobs et al., 2001; Janzen et al., 2004).

We first used a modified version of the water maze task where the platform is moved to a different location for each session. The goal was to assess if rats could rapidly memorize a new location in every session and improve their performance within sessions (between trial 1 and trials 2-3). During this task, the average escape latency and distance travelled by Dcx-KD rats decreased between trial 1 and trials 2-3, confirming Dcx-KD rats ability to learn properly (Figure 4A-B; repeated-measure 2-way ANOVA, trial effect, $F_{2,18}$ = 138.6, p < 0.0001 and $F_{2,18}$ = 86.81, p < 0.0001, respectively, n = 6). Moreover, their performance was not significantly different from that of mismatch rats demonstrating that short-term memory of Dcx-KD rats is not impaired in this context (repeated-measure 2-way ANOVA, trial x group interaction, $F_{2,18}$ = 1.371, p = 0.2791, n = 5-6 / condition).

We then used an 8-arm radial maze to further test working memory in Dcx-KD rats (Figure 4C-E). In the course of the 20 sessions, the number of arms unvisited within the allocated time (omissions) by Dcx-KD rats decreased while the number of correct choices increased, confirming once more that Dcx-KD rats can learn properly (Figure 4C-D; repeated measure 2-way ANOVA, session effect, $F_{4,36}$ = 32.29, p < 0.0001 and $F_{4,36}$ = 4.796, p = 0.0033, respectively, n = 6). Their performance was similar to the performance of mismatch rats, including their strategy to solve the task reflected by the mean angle between two consecutive visited arms (Figure 4C-E; repeated measure 2-way ANOVA, session x group interaction, $F_{4,36}$ = 0.066, p = 0.9917 (omissions); $F_{4,36}$ = 1.182, p =0.3351 (correct choices); $F_{4,36}$ = 1.149, p = 0.3492 (mean angle), n= 5-6 / condition).

Finally, we performed an object recognition task (Figure 4F). For both groups, time spent exploring objects decreased across sessions 2-7, indicating that rats became familiar with their environment, then rose when changes occurred (Figure 4G). Dcx-KD rats were able to detect the spatial change similarly to mismatch rats, as shown by re-exploration scores in Figure 4H (Displaced object: 21.56 ± 5.09 vs 23.13 ± 6.41; Non-displaced objects: -2.32 ±2.63 vs 3.78 ± 3.09; ordinary 2-way ANOVA, object x group interaction, $F_{1,18}$ = 0.232, p = 0.6358, n = 5-6 / condition). In addition, Dcx-KD rats detected the non-spatial change when a familiar object was replaced with a new one (Figure 4I; 37.88 ± 9.24 vs 29.82 ± 6.44; Mann-Whitney test, p = 0.5281, n= 5-6 / condition). Data gathered from these 3 tests indicate that short-term memory, including working memory and object recognition, functions normally in Dcx-KD rats.

## 3. DISCUSSION

In this study, we used a wide range of behavioral tests to assess anxiety, motor functions, somatosensory processing, learning and memory in a rat model of SBH. Our findings show that Dcx-KD rats do not display any obvious deficits in these processes. Nevertheless, these results do not preclude the existence of subtle alterations that could have gone undetected in our experiments. Similarly, other behaviors not investigated here, such as social interaction or competitive dominance, could be affected. In fact, competitive dominance was found to be lowered in mice models of SBH induced by the overexpression of Reelin or a constitutive-active form of Fyn (CA-Fyn), or a dominant-negative form of Cdk5 (Ishii et al., 2015).

It is important to emphasize that patients with SBH have a variable clinical picture, ranging from mild to severe deficits (Bahi-Buisson et al., 2013; Barkovich et al., 1994). Although some patients may have normal IQs, the majority of SBH patients have mild to moderate intellectual disabilities (23 out of 27 patients in the Barkovich et al. study, 1994, 76 out of 78 women with SBH in the study of Bahi-Buisson et al 2013). The absence of an apparent cognitive-behavioral phenotype in the young Dcx-KD rats studied here may be due to different factors, including the absence of epilepsy at the time of our analysis (Lapray et al., 2010), the size of SBH and the limitation of lissencephalic rodents brain to replicate human cortical phenotypes.

Epilepsy and intellectual disability are recurrent comorbidities that occur together and it has been a long-standing issue to understand how the former can affect the latter (see Holmes, 2015 and Lenck-Santini and Scott, 2015 for review). Although reported cognitive impairment in patients with epilepsy may be related to the underlying etiology (the heterotopia itself), there is strong evidence that seizures exacerbate cognitive outcomes and that antiepileptic drugs may contribute to cognitive impairment (Hermann et al., 2010; Kleen et al., 2012; Nalbantoglu et al., 2014). In addition, the severity of intellectual disability is strongly correlated with the age of epilepsy onset, so that patients with more severe cognitive deficits experienced their first seizure earlier (Barkovich et al., 1994; Tanaka and Gleeson, 2007). These later studies also indicate that the degree of cognitive impairment was also correlated with the anomalies of the cortex overlying the SBH: pachygyria and impaired gyri development. Thicker SBH and pachygyria were associated with more severe behavioral disturbances, polymorphic epileptic seizures, earlier onset of epilepsy, and resistance to antiepileptic drugs. Thus, it may be proposed that Dcx-KD rats do not develop a behavioral phenotype because rodents are lissencephalic or because cortical layers were essentially unaffected. Interestingly, the HeCO mouse, which display very large bilateral SBH following the mutation of the Eml1 gene (Kielar et al., 2014) and which also associates cortical atrophy, presents behavioral and cognitive alterations: approximately 40% of HeCo mice have a slow locomotor development and older HeCo mice exhibit cognitive deficits in a Morris water-maze paradigm (Croquelois et al., 2009).

It is important to emphasize that the majority of SBH mouse models that associate cognitive or behavioral deficits are models that essentially affect the cytoarchitectonic organization of the hippocampus (e.g. Dcx knockout mice (Corbo et al., 2002) and mice harbouring S140G mutation in Tuba1a (Keays et al., 2007)) and most of the observed deficits in these mice were identified in hippocampus-dependent tasks. It should be noted however that there is very few histopathological data describing dysplastic lesions of the human hippocampus though it has been suggested by Kuchukhidze et al. (2010), on the basis of MRI analysis, that one-third of patients with cortical dysplasia would also have an alteration of the hippocampus, mainly hypoplastic.

In conclusion, we propose that band heterotopia is not sufficient to significantly disrupt cognition and behavior and we suggest that other factors, such as gyral alterations and epilepsy, are required. While SBH models in rodents, such as Dcx-KD rats, are suitable for studying the pathophysiology of epilepsy (Lapray et al., 2010; Petit et al., 2014), higher mammals with gyrified cortex would be needed to study cognitive dysfunctions in cortical malformations.

## 4. EXPERIMENTAL PROCEDURES

### 4.1. ANIMALS

Animal experiments were performed in agreement with European directive 2010/63/UE.

#### *4.1.1. In utero electroporation*

Both mismatch controls and Dcx-KD animals were generated by in utero electroporation at embryonic day 15 (E15) as described previously (Martineau et al., 2018). Electroporated plasmids encoded either a shRNA targeting the 3'UTR of Dcx (mU6pro-3'UTRhp, gift from J. Loturco; Bai et al., 2003; 1 µg/µl) to generate Dcx-KD animals or an ineffective shRNA with 3 point mutations creating mismatches (mU6pro-3'UTR-mismatch, gift from J. Loturco; Bai et al,. 2003; 1 µg/µl) to create mismatch controls. A second plasmid encoding the green fluorescent protein (pCAG-GFP, Addgene #11150; 0.5 µg/µl) was co-injected as a reporter gene.

After birth, animals were screened for green fluorescence with specific goggles combining a light source emitting at 460-495 nm and a filter with an opening at 500-550 nm (BLS Ltd).

#### *4.1.2. Housing conditions*

Electroporated male Wistar rats were housed in groups of 2 littermates with a 12 h: 12 h light: dark cycle and dusk at 07:30. Food was provided *ad libitum* for all rats at least until the beginning of testing at 2 months old. At this point rats weighed between 300 g and 400 g. Afterward, weight gain was monitored for rats participating in the elevated plus maze, rotarod, beam walking, Morris water maze, radial maze and object exploration tasks and food intake was restricted to 10-20 g per rat daily. Water was always available in home cages.

### 4.2. BEHAVIORAL PROCEDURES

All rats were handled twice daily for 5 minutes during the 2 weeks preceding the experiments to ease up their stress. The behavioral tasks were performed in a dimly-lit familiar room, except when explicitly stated otherwise below. Testing occurred during the day portion of the circadian cycle (07:30-19:30) and both groups were always submitted to the different tasks at the same time of day. After each trial, dry mazes were cleaned with ethanol to neutralize olfactory cues that may bias the following trials. All tests except the rotarod, 8-arm radial maze and Von Frey hair test were filmed with an overhead black and white camera (either Sony SPT-M108 or DMK 23UP1300, Elvitec) at 25-30 frames per second. In tasks in which rats risked falling, a soft padded surface was placed at the base of the apparatus to cushion their fall.

#### 4.2.1. Anxiety and motor functions

*4.2.1.1. Elevated plus maze*

Anxiety was evaluated with the elevated plus maze (EPM) task. The apparatus was composed of a 70-cm high maze (Viewpoint, Champagne-au-Mont-d'Or, France) with four 50 cm long and 10 cm wide arms (two open arms and two closed arms with 70 cm high walls) that form a plus shape. Animals were placed in the center of the maze with their head facing an open arm and allowed to move freely during 5 minutes. To help videotracking, the maze was placed on top of an infrared floor (Viewpoint, Champagne-au-Mont-d'Or, France) consisting of a translucent floor under which a matrix of infrared lamp was installed to help increase the contrast between the background and animal.

*4.2.1.2. Rotarod*

Motor coordination was first tested on a rotating beam (Rotarod LE8305; Bioseb). Rats performed 10 trials a day during 3 consecutive days, with a resting period of about 5 minutes between trials. Each day, speed was increased by 5 revolutions per minute (rpm), starting at 5 rpm on day 1 and ending at 15 rpm on day 3. We monitored the latency to fall with a cut-off at 3 minutes. Results are the mean of 10 daily trials.

*4.2.1.3. Beam walking*

Motor coordination was further assessed by having rats walk on a 2 cm-wide and 1 m-long horizontal bar placed 45 cm above the floor. Both ends of the beam were blocked with cardboards so that rats could not escape. Rats were held onto the middle of the horizontal bar until they found their balance then allowed to move freely. Two trials held 45 minutes apart were recorded; the first one was used to familiarize the animals with the apparatus, the second one was analyzed. A trial ended when the rat fell or after 3 minutes.

### 4.2.2. Somatosensory functions

*4.2.2.1. Von Frey hair test*

Tactile sensitivity of the hind paws was evaluated with the Von Frey hair test (Chaplan et al., 1994; Sánchez-Brualla et al., 2017). The up and down method was used to establish the 50% threshold as described previously (Chaplan et al., 1994; Gaillard et al., 2014). Briefly, the first stimulation was made with the 8 g filament (Von Frey filaments kit, Bioseb) and response to this stimulation was assessed. If the response was positive, the $2^{nd}$ stimulation was made with a thinner filament (6g), otherwise it was made with a thicker filament (10 g). Following this rule, stimulations continued until a change in the response of the animal was recorded. After this point, 4 final up and down stimulations were performed.

*4.2.2.2. Adhesive removal*

The adhesive removal test allowed to assess the haptic sense of the front limbs. Small round adhesive-backed labels were attached to the front limbs of rats on each side. Each front paw received a small adhesive (8 mm diameter; Scotch) on their dorsal part and a slightly larger adhesive (12 mm dimeter; Avery Zweckform) was placed on the middle of each forelimb. All 4 adhesives were attached in a random order by the experimenter. Rats were then placed in a familiar cage until they removed all 4 labels, with a cut-off time set at 2 minutes. This task was replicated 4 times for each rat with a resting period of about 2 minutes between successive trials. Two cameras were used to simultaneously film each side of the translucent cage and catch all of the rat movements.

*4.2.2.3. Gap crossing*

The gap crossing test was used to check vibrissae-dependent object perception. The training apparatus consisted of two platforms (10x19 cm, 15 cm high), one of which was closed on 3 sides and on top (15x10x12 cm) to create a shelter. The experiment was conducted in a room lit with a bright light to reinforce the natural need for rats to seek shelter. In addition, some litter from the home cage was placed in the closed platform to identify it as a safe space. At the beginning of the experiment, rats were placed on the open platform and no gap was left between the two platforms. Each time rats crossed, they were allowed to rest in the shelter for 1 minute then were placed back onto the open platform with a 1 cm gap increase. Rats were given 2 trials per gap distance with a cut-off time set at 2 minutes. If by that time, rats had not crossed over, the maximal distance crossed was recorded and the test ended.

This task was performed a second time by all animals after their left vibrissae were clipped with scissors. Rats were allowed a resting period of 45 minutes between the clipping and the test.

*4.2.2.4. Texture discrimination*

The apparatus consisted of a black plastic box with an open top, divided into 2 identical square compartments (30x30x40 cm) and a corridor in-between them (10x30x40 cm). The compartments were connected to the corridor through two openings (10x15 cm) that could be closed using removable doors. Black curtains surrounded the box to block out spatial cues. The test was divided in 2 parts: the 1st part ensured that rats were responsive to aversive conditioning, the 2nd one tested fine discrimination of textures. Each part of the test was divided into 3 sessions of 4 minutes with a 4 minute-intersession interval during which rats were returned to a familiar cage (see Figure 2F). During session 1, rats were placed in the corridor and allowed to move freely to become familiar with the testing environment. If rats displayed a clear preference for one of the compartments, it was chosen as the devaluated compartment for session 2; otherwise, the devaluated compartment was assigned at random. In session 2, rats were placed in the chosen compartment with the door closed. A strobe light (IBIZA strobe light DMX, Electronic Star) placed on the table behind the box was manually set on during 20 seconds every 40 seconds (1500 W, 13 Hz). Finally, in session 3, the door was removed, rats were placed in the corridor with their head facing a different direction than in session 1 and allowed to move freely. During part 1, the two compartments were made distinguishable by covering the walls of one with black sheets and of the other with white sheets. In part 2, the light was switched off and replaced with a dim red light (25W) placed outside the curtains. Infrared LED (SOLAROD LED 850 nm) lit the box which had been rotated by 90 degree to disorient the animals. The colored sheets were stripped off the walls and replaced with sandpaper (Weldom, Marseille, France). One compartment held fine (P180) sandpaper and the other one coarse (P40) sandpaper, including on the floor. Sessions 4 to 6 were identical to sessions 1-3 but were carried out in this modified environment. Throughout the test, all sheets of colored paper and sandpaper were changed between the 2nd and 3rd sessions to remove potential olfactory cues.

#### 4.2.3. Learning and memory

*4.2.3.1. Morris water maze*

Learning and memory were tested using Morris water maze (MWM) as described in Van Cauter et al., 2013 with the following modifications. Training was carried out during 4 days with 2 daily sessions of 4 trials and a 30-second inter-trial interval during which the rat was left on the platform and a between-session interval of 4 hours minimum. Once training was complete, the animals underwent two probe trials consisting of 1 minute spent in the pool without the platform. The first probe test was carried out 24 hours after completion of training, the second one, 10 days later.

In a second phase, rats were trained using a different protocol in which the platform was moved across sessions. This protocol comprised 2 daily sessions of 3 trials during 4 days. On each trial, the rat was started from a randomly changing starting point. Inter-trial time lasted 2.5 minutes including 30 seconds spent on the platform. The platform was located 30 or 45 cm away from the wall so that the total 8 positions used formed 2 squares, with the inner square tilted 45 degree in relation to the outer square.

### 4.2.3.2. *Social transmission of food preference (STFP)*

Remote associative olfactory memory was probed with a social transmission of food preference paradigm carried out as previously described (Bessières et al., 2017). Briefly, 10 non-electroporated Wistar rats (3-4 months old), thereafter called "demonstrators", were habituated to cumin-flavored food (powdered food with 0.5% cumin, 15 g available every day for 30 minutes) for 3 days. On the 4th day, demonstrators interacted with a mismatch or Dcx-KD rat ("observer") for 30 minutes so that observers could smell cumin on their breath. From days 30 to 33, observers were taught to feed on plain powdered food (Safe Diets) that was made available for only 20 minutes each day. The preference test took place on day 34 in the home cage of observers. Powdered food was mixed with either cumin (0.5%) or thyme (0.75%). Both types of food were presented to the observers for 30 minutes after which the dishes were removed and the remaining food weighted. Because observers smelled cumin on the demonstrator's breath, this food is supposed to be considered safe and preferentially eaten by observers when exposed to it one month later.

### 4.2.3.3. *8-arm radial maze*

Working memory was tested using an 8-arm radial maze task. The apparatus was an elevated (50 cm high) maze with 8 arms (50 cm long, 10 cm wide) radiating from an octagonal shaped central platform (20 cm in diameter). A food cup (5 cm diameter) was placed 1 cm from the distal part of each arm. Numerous extra maze cues were visible in the room. During testing, food was restricted to 10 g daily per rat and weight was monitored to ensure it would not go down below 85% of the free-feeding weight. Two days before the experiment, approximatively 5 g of chocolate cereals Cocopops (Kellog's) were placed in each cage to familiarize the animals with the taste. A typical training trial consisted of having an animal placed on the central platform and allowed to visit all 8 baited arms to eat the Cocopops (1/arm) with a time limit of 3 minutes. Rats were trained 2-4 times daily for 8 days amounting to a total of 20 training sessions. Rats of both groups received the same amount of daily training time and 2 sessions were spaced out by at least 1 hour.

*4.2.3.4. Object recognition*

Object recognition was carried out the experiment as previously described (Save et al., 1992; Van Cauter et al., 2013). Briefly, the experiment consisted of ten 4-minute sessions separated by 4-minute intervals during which animals returned to their home cage (see Figure 4F). During the first session, the animal was familiarized with the environment by being placed in the arena without any object. In sessions 2 to 7, 4 objects were present in the arena and remained in the same position all throughout these 6 sessions. Before the 8$^{th}$ session, one object was displaced about 25 cm away from its previous location ("spatial change"). Finally, in session 10, one object was replaced with a novel object ("non-spatial change"). According to previous reports, these changes normally trigger a selective re-exploration of the objects (Save et al., 1992; Van Cauter et al., 2013).

**4.3. DATA ANALYSIS**

*4.3.1. EPM, beam walking and MWM*

A videotracking software (Videotrack, Viewpoint, Champagne-au-Mont-d'Or, France) was used to record and analyze the rats' behavior during the EPM, beam walking and MWM tasks. In all trials, the trajectory length, duration and moving speed were measured. For the probe trials of the MWM task, we additionally calculated an additional error score (E), which reflects the dispersion of searching behavior independently from the distance swum by the animal (Granon and Poucet, 1995) and is defined as:

$$E = \sqrt{\sum_{i=0}^{n} d^2 / 2n}$$ where *d* is the distance to the goal at each point of the rat's swimming path and *n* is the total number of points of the path.

*4.3.2. Objects exploration, adhesive removal and texture discrimination*

Tasks were recorded with a video camera and analyzed off line. The annotation software Anvil (Kipp, 2012) was used to analyze both the object exploration and texture discrimination tasks. For object exploration, the number and duration of explorations were analyzed. An object exploration was counted when the snout was directed at a distance of less than 2 cm of an object or when the rat actively contacted the object with its snout or its paw. The re-exploration scores (RS) were calculated as follows: for the spatial change $RS = T_{S8} - T_{S7}$ where $T_{S8}$ and $T_{S7}$ are times spent exploring the object(s) during session 8 and session 7 respectively (for the non-displaced objects, T is the average of time spent exploring the 3 objects) ; for the non-spatial change

$$RS = T_{NO} - (T_{FO1} + T_{FO2} + T_{FO3})/3$$ where $T_{NO}$ and $T_{FO}$ are times spent exploring the novel and familiar objects during session 9, respectively (Van Cauter et al., 2013).

For the texture discrimination task, an entry was counted when all four paws of the animal were in the compartment. Duration and number of entries were recorded for each compartment in each session. The discrimination score was defined as

$$DS = (T_{S3} - T_{S1}) * 100 / T_{S1}$$ where *T* is the time spent in compartment and *S* is session.

For adhesive removal, the two videos of each trial were synchronized in Vegas Pro 13 (Sony). Time to contact and remove each adhesive label was thus measured on the software, going back and forth between both tracks to counterbalance the rats' quick movements.

*4.3.3. Rotarod, STFP, 8-arm radial maze, Von Frey hair test and gap crossing*

For the Von Frey hair test, the 50% threshold was calculated using a custom python program following Chaplan *et al.* method (Chaplan et al., 1994). For all other tests, relevant data described in the results section were manually recorded at the time of testing. For the 8-arm radial maze, an arm entry was counted when rats placed all 4 paws inside the arm.

### 4.4. STATISTICS

All statistical analyses were performed using Graphpad Prism 6. Comparison of groups was tested with Mann-Whitney test, ordinary two-way ANOVA, repeated measure two-way ANOVA or binomial tests. All values are given as mean ± SEM. All tests were two-tailed and the level of significance was set at $p < 0.05$. Statistical power was checked for significant analyses.

**AUTHOR CONTRIBUTIONS**

F.S.M. and L.F. carried out behavioral procedures. F.S.M., F.S and B.P. analyzed the data. E.B. performed *in utero* electroporations. L.F. and F.W. prepared the plasmids. F.S.M., F.W., F.S., JB.M., B.P. and A. R. designed and conceived the experiments and wrote the article.

**FIGURE LEGENDS**

Figure 1: Motor functions are preserved in Dcx-KD rats

(A) Images of mismatch and Dcx-KD brain slices taken in transmission light. The two red arrows indicate the presence of a SBH in the Dcx-KD brain. Right panels show a higher magnification of the left panels. Scale bars: left = 2 mm, right = 1 mm

(B, C) Percentage of time spent in the open arms (B) and locomotor activity (C) during the elevated plus maze task (EPM)

(D) Latency to fall from the rotarod at 5, 10 and 15 rpm

(E-G) Latency to fall (E), path length (F) and speed (G) during the beam walking test

Figure 2: Somatosensory functions of Dcx-KD rats are unaffected by the SBH

(A) Mechanical 50% threshold for the left and right hind paws calculated from the Von Frey hair experiment

(B, C) Latency to contact dots on the left and right front limb during the adhesive removal test (B). Percentage of rats which first contacted a dot on their left / right side during their first trial (C)

(D, E) Maximal distance crossed during the gap crossing test with intact (D) or clipped left (E) vibrissae

(F-H) Schematic representation of the texture discrimination task protocol (F). The lightning figures the aversive conditioning carried out during sessions 2 and 5 using the strobe light. Discrimination scores of the devaluated and non-devaluated compartment during sessions 3 (G) and 6 (H)

Figure 3: Learning and long-term memory are essentially normal in Dcx-KD rats

(A) Representative traces of paths taken by rats to reach the platform at the beginning of training for Morris water maze, mid-training and at the end of training

(B, C) Escape latencies (B) and path lengths (C) across training sessions for Morris water maze

(D-F) Time spent swimming (D) and distance swam (E) in the target and opposite quadrants during a probe test carried out 24 hours after the last training session for Morris water maze. The error score is illustrated in (F)

(G-I) Time spent swimming (G) and distance swam (H) in the target and opposite quadrants during a probe test carried out 10 days after the last training session for Morris water maze. The error score is illustrated in (I)

(J, K) Timeline of the social transmission of food preference experiment (J). D: demonstrator; O: observer; d: day. Percentage of cumin- and thyme-flavored food eaten by rats during the preference test (K)

Figure 4: Short-term memory, including working memory, functions normally in Dcx-KD rats

(A, B) Escape latencies (A) and path lengths (B) across trials of a modified version of Morris water maze in which the platform is in a new location every day

(C-E) Number of arms unvisited during the allocated time (omissions) (C), number of correct choices among the first 8 arm entries (D) and mean angle between two consecutive arm entries (E) across blocks of 4 sessions in the 8-arm radial maze (F-I) Schematic representation of the object recognition task protocol (F) and total time spent exploring objects across sessions 2-10 (G). Re-exploration scores of the displaced and non-displaced objects during the spatial change session (H) and of the novel object during the last session (I).